\documentclass{iau}
\usepackage{graphicx}

\newcommand{\eq}[1]{\begin{equation}  #1 \end{equation}}

\newcommand{\ba}[1]{\left\langle #1 \right\rangle}

\newcommand{\vek}[1]{\mbox{\boldmath $#1$}}

\title[SCCC21.~~Errors on errors] 
{Errors on errors --\\ Estimating cosmological parameter covariance}

\author[Benjamin Joachimi \& Andy Taylor]   
{Benjamin Joachimi$^1$ \and Andy Taylor$^2$}

\affiliation{$^1$Department of Physics \& Astronomy, University College London, \\ Gower Place, London WC1E 6BT, United Kingdom\\ email: {\tt b.joachimi@ucl.ac.uk} \\[\affilskip]
$^2$Institute for Astronomy, University of Edinburgh, \\ Royal Observatory, Blackford Hill, Edinburgh EH9 3HJ, United Kingdom \\email: {\tt ant@roe.ac.uk}}

\pubyear{2014}
\volume{306}  
\pagerange{?--?}
\jname{Statistical Challenges in 21st Century Cosmology}
\editors{A.F. Heavens, J.-L. Starck, A. Krone-Martins eds.}
\begin{document}

\maketitle

\begin{abstract}
Current and forthcoming cosmological data analyses share the challenge of huge datasets alongside increasingly tight requirements on the precision and accuracy of extracted cosmological parameters. The community is becoming increasingly aware that these requirements not only apply to the central values of parameters but, equally important, also to the error bars. Due to non-linear effects in the astrophysics, the instrument, and the analysis pipeline, data covariance matrices are usually not well known a priori and need to be estimated from the data itself, or from suites of large simulations. In either case, the finite number of realisations available to determine data covariances introduces significant biases and additional variance in the errors on cosmological parameters in a standard likelihood analysis. Here, we review recent work on quantifying these biases and additional variances and discuss approaches to remedy these effects.
\end{abstract}

\firstsection 
\section{What noise does to your covariance matrix}

The variance-covariance of cosmological data is often complicated and not known analytically, receiving contributions from sample variance coupled with complex survey masks, instrumental effects, as well as measurement and shot noise. Therefore it is customary to estimate data covariance matrices from the data itself (via resampling techniques) or from suites of realistic mock datasets. The latter are computationally expensive to generate, so that there is a strong drive to keep the number of simulated realisations of the data, $N_S$, to a minimum. Conversely, the ever increasing size of surveys and number of mature cosmological probes to be extracted leads to data vectors which can easily exceed dimensions of $N_D > 1000$ in the near future.

Despite the pressure to keep $N_S$ small while creating data vectors with large $N_D$, cosmologists have until recently almost exclusively employed the standard sample covariance estimator and proceeded to perform a likelihood analysis without further consideration of the statistical uncertainty and potential biases of their covariance estimate. Only recently, beginning with the work of \cite[Hartlap et al. (2007)]{hartlap07}, has there been an increased awareness of long-established results on this topic in the statistics literature.

\begin{figure}[t]
\begin{center}
 \includegraphics[width=4cm,angle=270]{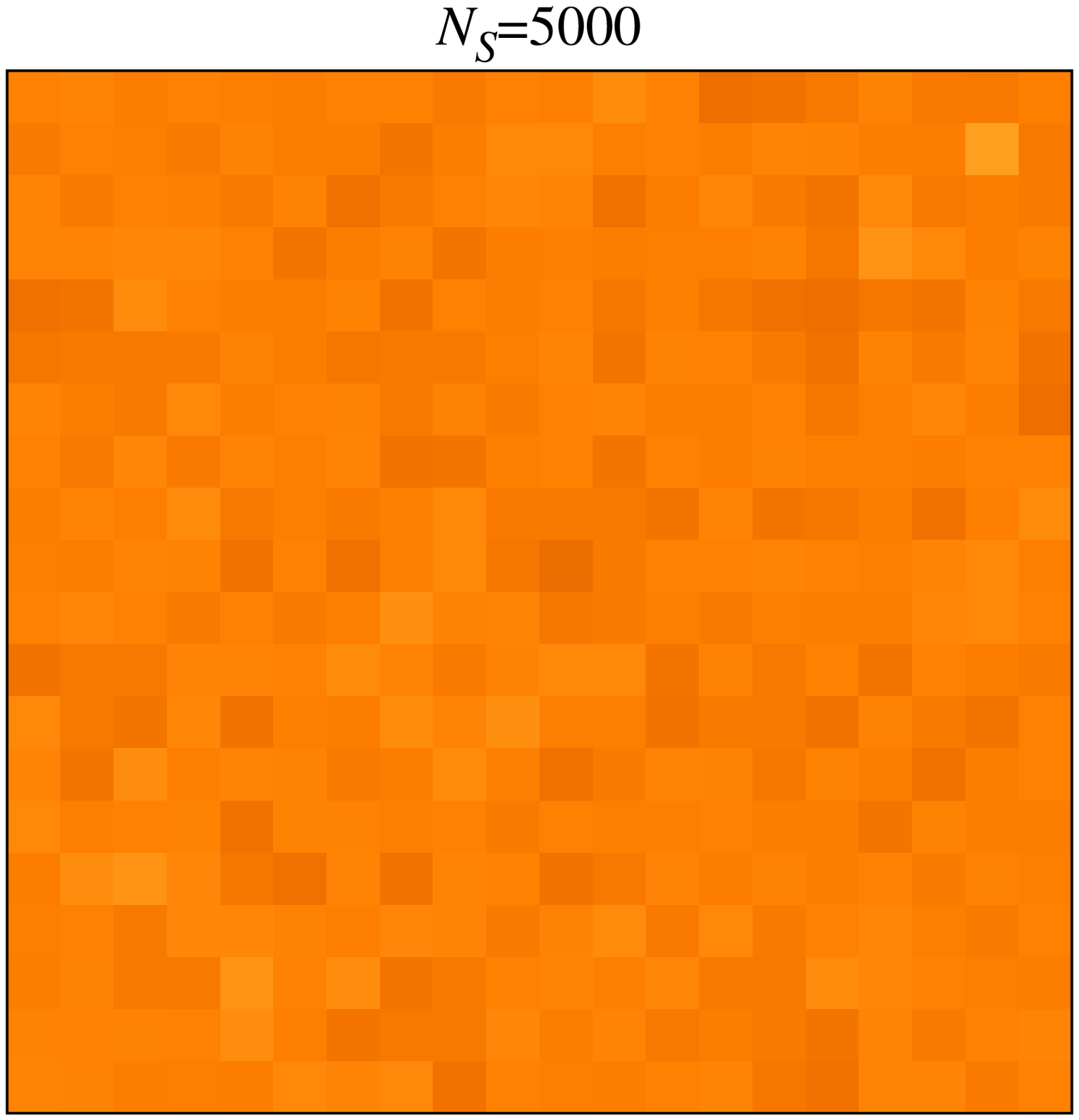}%
 \includegraphics[width=4cm,angle=270]{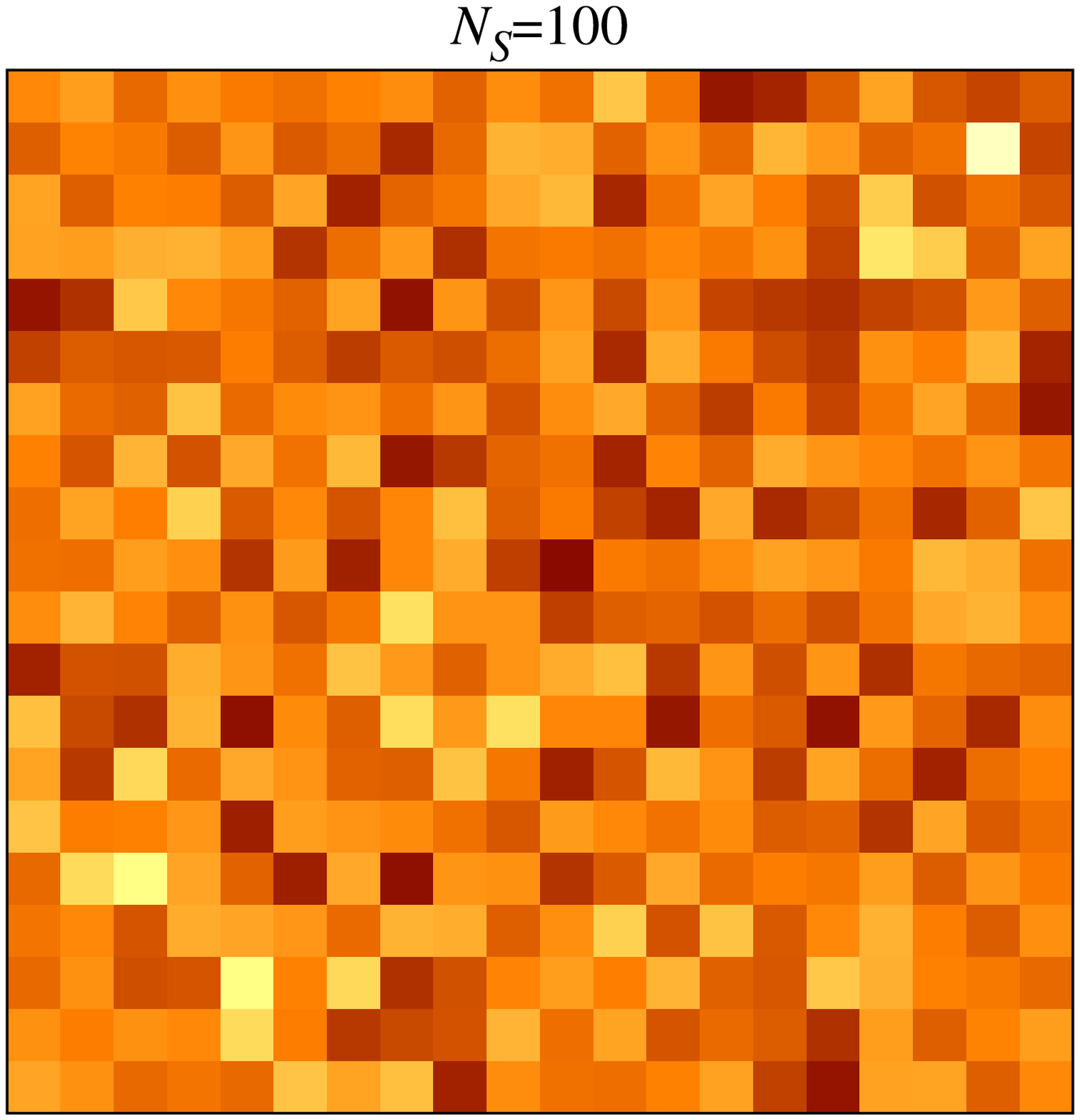}%
 \includegraphics[width=4cm,angle=270]{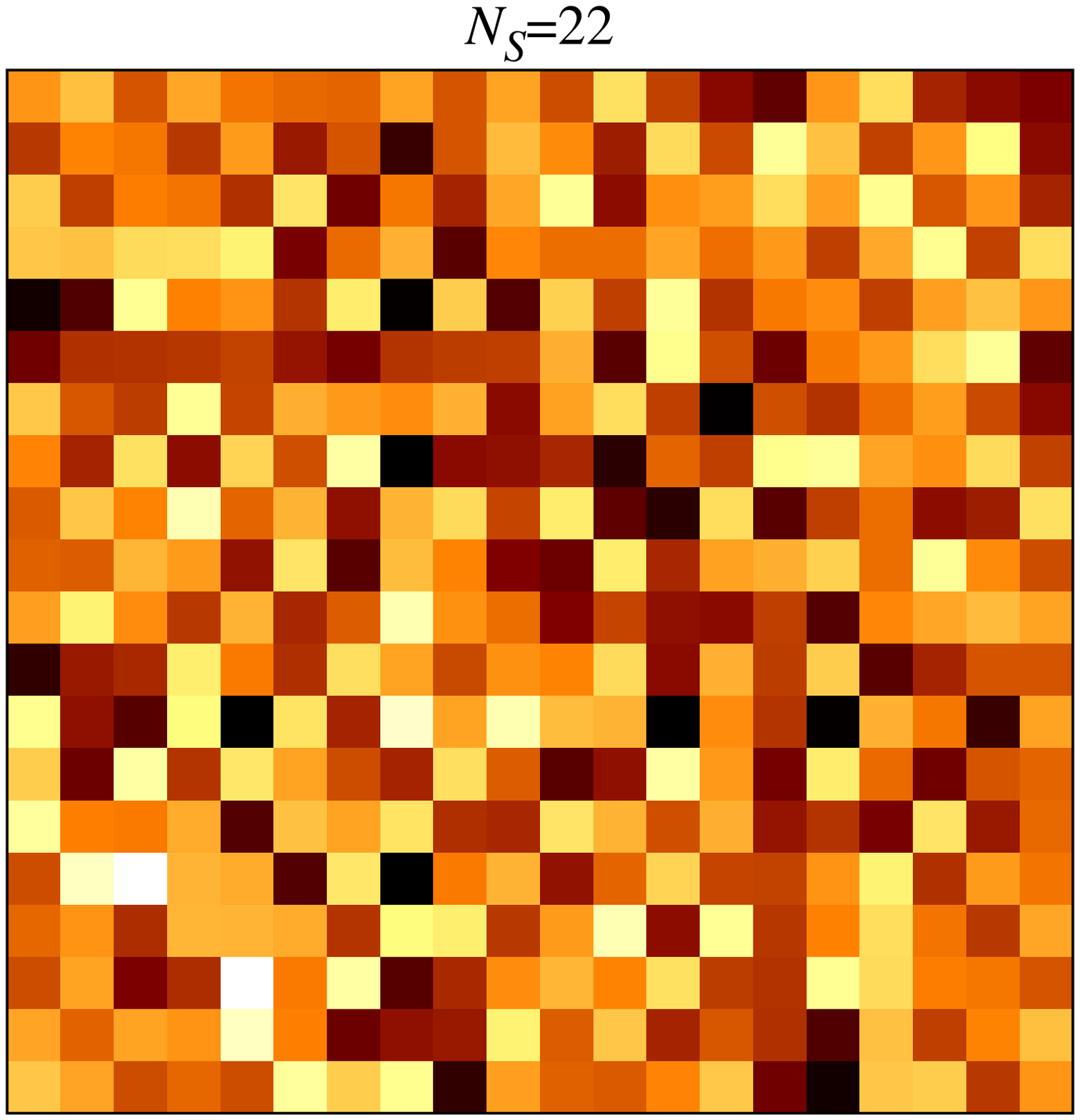}\\
   \includegraphics[width=4.5cm,angle=270]{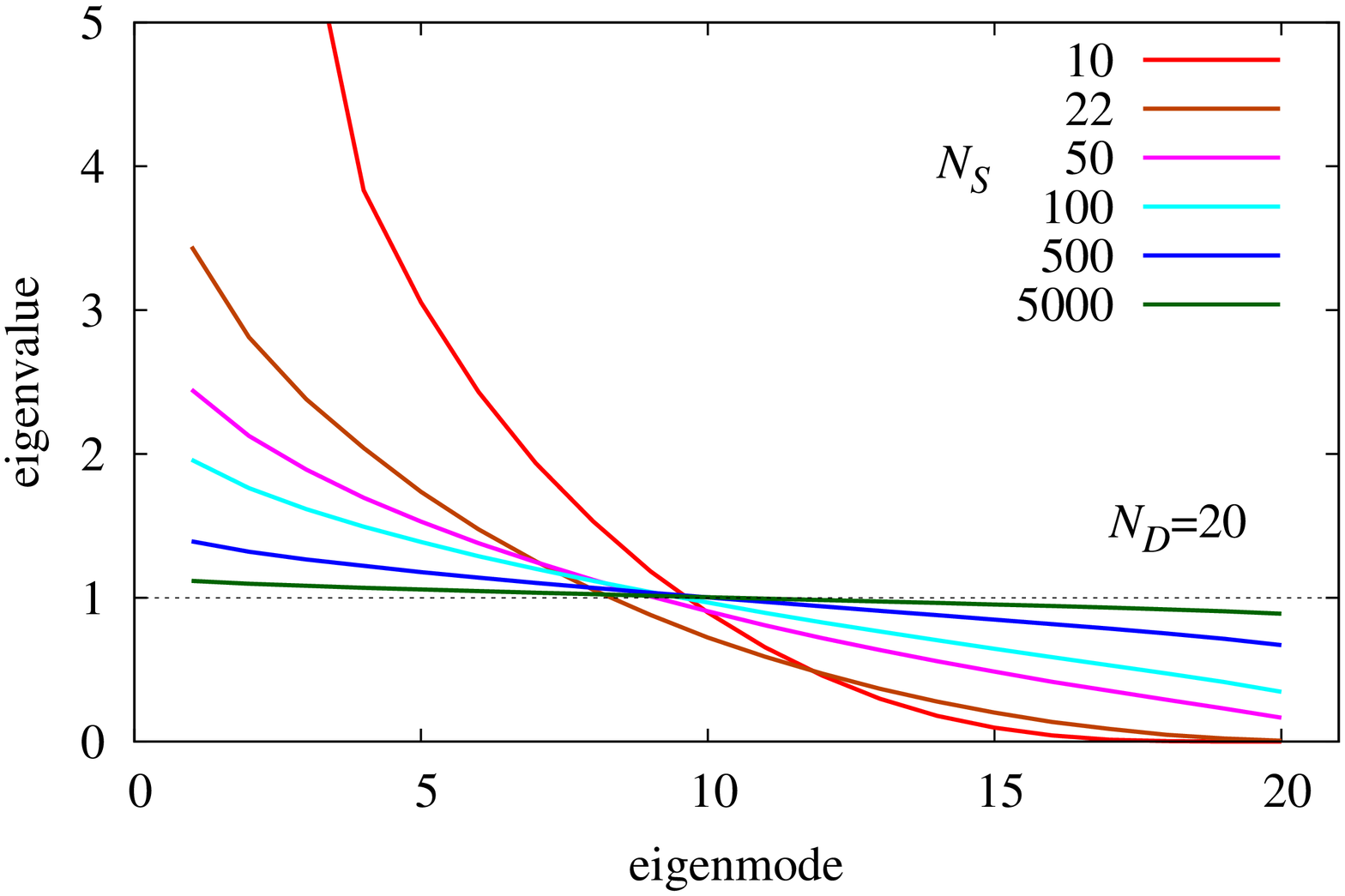} 
 \caption{Illustration of the impact of noise on a covariance matrix, for the toy case of a 20-dimensional identity matrix. \textit{Top}: Realisations of Wishart-distributed random matrices with expectation subtracted, generated for $N_S=5000,100,22$ (from left to right) to illustrate the increasing levels of noise. \textit{Bottom}: Ordered eigenvalues, averaged over 1000 realisations of covariance matrices. The lines show eigenvalues for different values of the number of realisation used for computing the covariance, $N_S$, as indicated in the legend. The dashed horizontal line indicates the eigenvalues in the noise-free case.}
   \label{fig1}
\end{center}
\end{figure}
 
If the elements of a data vector $\vek{D}$ are Gaussian distributed, the data sample covariance estimate $\mathbf{M}=\ba{\Delta \vek{D}\; \Delta \vek{D}^\tau}$, where $\Delta \vek{D}=\vek{D} - \ba{\vek{D}}$, follows a Wishart distribution (the multivariate generalisation of a $\chi^2$ distribution) with $N_S-1$ degrees of freedom \cite[(Wishart 1928)]{wishart28}. The inverse data covariance, $\mathbf{\Psi} \equiv \mathbf{M}^{-1}$, which is required in least squares and likelihood expressions, then follows an Inverse-Wishart distribution with $N_S-N_D-2$ degrees of freedom. The moments of this distribution were first derived by \cite[Kaufman (1967)]{kaufman67} who found for the mean
\eq{
\label{eq:inv}
\ba{\mathbf{\hat{\Psi}}} = \frac{N_S-1}{N_S-N_D-2} \mathbf{\Psi}\;,
}
demonstrating that $\mathbf{\hat{\Psi}}$ is increasingly biased high for decreasing $N_S$ and diverges at $N_S=N_D+2$. Figure \ref{fig1} provides an illustration for this non-intuitive behaviour. For decreasing $N_S$ the largest (smallest) eigenvalues of a noisy covariance matrix are biased increasingly high (low), and the condition number dramatically increases. The smallest eigenvalue drops to zero at $N_S=N_D+2$, rendering the covariance singular. Even after correcting for the bias, the variance in the covariance estimate diverges at a very similar rate \cite[(see again Kaufman 1967)]{kaufman67}.

\begin{figure}[t]
\begin{center}
 \includegraphics[width=9cm]{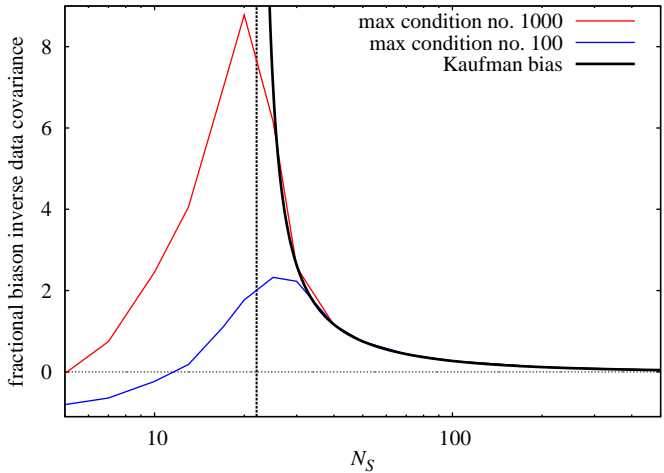} 
 \caption{The effect of noise stabilising measures (via singular value decomposition) on the bias of the inverse covariance. Shown is the average fractional bias on the diagonal elements of the inverse covariance matrix (for $N_D=24$; indicated by the vertical line), as a function of the number of realisation used for computing the covariance, $N_S$. The black solid line corresponds to the bias of the inverse of the sample covariance matrix, as calculated by \cite[Kaufman (1967)]{kaufman67}. The red (blue) line shows the bias for the case that the smallest eigenvalues of the covariance have been set to zero before calculating a pseudo-inverse, such that the condition number does not exceed 1000 (100).}
   \label{fig2}
\end{center}
\end{figure}
 
One may wonder about the impact on early cosmological analyses which generally boasted only a small number of simulations and seem to have ignored the biases in the inverse data covariance. Since the estimated covariances would have been very noisy, one can hypothesise that researchers would have calculated a singular value decomposition and set the smallest eigenvalues to zero before proceeding with a Moore-Penrose pseudo-inverse. Figure 2 illustrates two possible scenarios, alongside with the bias expected from Eq. (\ref{eq:inv}). Depending on the exact value of $N_S$ and $N_D$, and the details of the noise-suppression measures, biases may have been large and both positive and negative. The increase in the largest eigenvalues would have remained untreated, and consequently the inverse covariance would have tended to zero (or the fractional bias to -1) for $N_S \rightarrow 0$.
 
Precise and accurate cosmological analysis requires a more careful treatment of these noise effects, either quantifying their effect on cosmological parameter errors or mitigating them by employing alternatives to the standard sample covariance estimator.

 \section{Impact on the errors of cosmological parameters}

\cite[Taylor \& Joachimi (2014, TJ14 hereafter)]{taylor14} calculated the impact of biases and variances in the data sample covariance on the cosmological parameter covariance, finding that the latter is generally biased high and takes up additional variance \cite[(see Dodelson \& Schneider 2013 and Percival et al. 2014 for earlier, approximate results)]{dodelson13,percival14}. For the case that the parameter covariance is estimated from the curvature of the likelihood at its peak (similar to estimates from MCMC samples), they derived its full distribution, which is again a Wishart distribution with $N_S-N_D+N_P-1$ degrees of freedom, where $N_P$ is the number of parameters, i.e. the dimension of the parameter covariance matrix. The mean reads
\eq{
\label{eq:cw}
\ba{\mathbf{\hat{C}}^W_{\mu \nu}} = \frac{N_S-N_D+N_P-1}{N_S-N_D-2}\; \mathbf{C}_{\mu \nu}\;,
}
while exact expression for the variance can also be derived. Additionally, TJ14 found an exact expression for the mean of the parameter covariance estimated from the scatter in likelihood peaks,
\eq{
\label{eq:cp}
\ba{\mathbf{\hat{C}}^P_{\mu \nu}} = \frac{N_S-2}{N_S-N_D+N_P-2}\; \mathbf{C}_{\mu \nu}\;,
}
which they determined from simulations based on random Wishart matrices. De-biasing the parameter covariances based on these expressions, TJ14 derived constraints on the minimum number of simulations required to reach a maximum contribution, $\nu$, to the error on cosmological parameters originating from noise in the data covariance (see their Fig. 5). In the case of the parameter covariance derived from the likelihood curvature, $\mathbf{C}^W$, the minimum number is given by $N_S \approx N_D + N_P/\nu +2 \nu^{-2}$. For forthcoming cosmological surveys, where easily $N_D \sim 1000$, this implies that, if one tolerates a $10\%$ increase in parameter errors due to noise in the data covariance, $N_S$ has to be only slightly larger than $N_D$, whereas if one restricts this increase to $1\%$, a challenging number of $N_S > 10^4$ would result.

\section{Beyond the sample covariance}

There are various alternatives to using the standard sample covariance estimator in a likelihood analysis, which \cite[Taylor et al. (2013, TJK13 hereafter)]{taylor13} presented in schematic form in their Fig. 8. Below we summarise the most relevant points.

{\underline{\textbf{Resampling methods}}: We have concentrated here on the generation of realisations of the data from simulations, i.e. externally. Resampling methods like bootstrap or jackknife allow for the estimation of covariances internally from the data. To preserve the correlations in the data, one has to define blocks of survey volume, or patches on the sky, which are quasi-independent. This requirement limits the number of blocks for a given survey volume/area. Moreover, long-range correlations will invalidate this assumption to some degree and cause biases. In addition, bootstrap and jackknife estimates are generally biased and, although consistent, can converge slowly. For more details on resampling methods see the contribution by P. Arnalte-Mur.

{\underline{\textbf{Data compression}}: Data compression alleviates the problem of noise effects originating from the data covariance as biases and variances scale approximately with $N_S-N_D$. Maximal data compression, i.e. $N_D \rightarrow N_P$, even eliminates all adverse noise effects, as can be seen from Eqs. (\ref{eq:cw}) and (\ref{eq:cp}). Note that in the latter case the factor from the Kaufman bias in Eq. (\ref{eq:inv}) remains, which can be removed as the parameter covariance is now a linear transformation of the data covariance, so no inversion is necessary. However, data compression techniques require some information about the data covariance. TJK13 show for maximal Karhunen-Loeve compression that, using the sample covariance in the compression operation, exactly reproduces the original noise effects -- nothing is gained. Employing a noise-free model covariance instead renders the compression suboptimal, thus increasing errors on cosmological parameters to a yet unknown degree.

{\underline{\textbf{Shrinkage}}: We are not completely ignorant about the form of covariance matrices for cosmological data. Their elements vary smoothly across angular and redshift scales, and sometimes it is fair to assume that the diagonal dominates the errors, e.g. if shot noise is prominent. In some regimes, such as on large, nearly Gaussian scales, we may even have good, if not exact, analytic models. This is valuable prior information that can be used to improve covariance estimates. One such approach is shrinkage estimation, building a linear combination of the sample covariance and a model covariance (which can contain free parameters). The weighting of the linear combination can be estimated analytically from the data. TJK13 test several shrinkage estimators of covariance for a toy cosmological case \cite[(see also Pope \& Szapudi 2008)]{pope08}. Their worst-performing estimator was derived by \cite[Stein et al. (1972)]{stein72} who proved that nonetheless their estimator outperforms the sample estimator with respect to a \lq natural\rq\ loss function based on the mean square error of the mean vector of the data, which implies the sample covariance estimator is inadmissible (see also D. van Dyk's contribution).

\end{document}